\RequirePackage{fix-cm}
\documentclass[twocolumn,epjc3]{svjour3} 
\usepackage{multirow}  
\usepackage[dvipsnames]{xcolor}

\usepackage[pagewise,mathlines]{lineno}

\smartqed  
\RequirePackage{graphicx}
\RequirePackage[numbers,sort&compress]{natbib}
\journalname{Eur. Phys. J. Plus}
\begin{document}

\title{Characterization of a novel proton-CT scanner \\based on Silicon and LaBr$_3$(Ce) detectors}



\author{
        E. N\'acher\thanksref{addr1, e1} 
        \and
        J.A. Briz\thanksref{addr2, e2}
        \and
        A.N. Nerio\thanksref{addr2}
        \and
        A. Perea,\thanksref{addr2}
        \and
        V.G. T\'avora\thanksref{addr2}
        \and
        \\O. Tengblad\thanksref{addr2}
        \and
        M. Ciemala\thanksref{addr3}
        \and
        N. Cieplicka-Orynczak\thanksref{addr3}
        \and
       A. Maj\thanksref{addr3}
        \and
        K. Mazurek\thanksref{addr3}
        \and
        P. Olko\thanksref{addr3}
        \and
        M. Zieblinski\thanksref{addr3}
        \and
        M.J.G. Borge\thanksref{addr2}
}

\thankstext{e1}{corresponding author: enrique.nacher@csic.es}
\thankstext{e2}{present address: Universidad Complutense de Madrid, CEI Moncloa, E-28040 Madrid, Spain}


\institute{Instituto de F\'\i{}sica Corpuscular, CSIC - Univ. de Valencia, E-46980 Paterna, Valencia, Spain \label{addr1}
           \and
           Instituto de Estructura de la Materia, CSIC, E-28006 Madrid, Spain \label{addr2}
           \and
           Instytut Fizyki Jadrowej PAN, 31-342 Krakow, Poland \label{addr3}
}

\date{Received: date / Accepted: date}

\maketitle

\begin{abstract}
\noindent Treatment planning systems at proton-therapy centres entirely use X-ray computed tomography (CT) as primary imaging technique to infer the proton treatment doses to tumour and healthy tissues. However, proton stopping powers in the body, as derived from X-ray images, suffer from important proton-range uncertainties. In order to reduce this uncertainty in range, one could use proton-CT images instead. The main goal of this work is to test the capabilities of a newly-developed proton-CT scanner, based on the use of a set of tracking detectors and a high energy resolution scintillator for the residual energy of the protons. Different custom-made phantoms were positioned at the field of view of the scanner and were irradiated with protons at the CCB proton-therapy center in Krakow. We  measured with the phantoms at different angles and produced sinograms that were used to obtain reconstructed images by Filtered Back-Projection. The obtained images were used to determine the capabilities of our scanner in terms of spatial resolution and proton Relative Stopping Power (RSP) mapping and validate its use as proton-CT scanner. The results show that the scanner can produce medium-high quality images, with spatial resolution better than  2 mm in radiography, below 3 mm in tomography and resolving power in the RSP comparable to other state of the art pCT scanners.

\end{abstract}

\section{Introduction}

According to World Health Organisation, cancer is the leading cause of death in the world. More than 50\% of cancer patients receive some kind of radiation therapy (radiotherapy) during their course of treatment. Conventional radiotherapy for deep tumours makes use of X rays to control or kill malignant cells. Unfortunately, healthy tissue is not immune to the ionisation produced by the X rays and, therefore, the areas surrounding the cancerous tumour are severely damaged. Proton therapy is a technique that uses proton beams instead of X rays as ionising radiation. It has a far higher dose selectivity than conventional radiotherapy, what makes it ideal for the treatment of localised tumours in highly sensitive areas e.g. brain, heart or spinal cord.

The application of proton therapy, however, is not exempt of difficulties. The precision in the determination of the distal position of the dose distribution is crucial for a complete irradiation of the tumour and to avoid, as much as possible, any dosage to the surrounding healthy tissue. So far, treatment planning systems at proton-therapy centres use X-ray computed tomography (X-ray CT) as primary imaging technique to calculate doses to tumour and healthy tissues. This produces a map of the linear attenuation coefficient of the tissue for X rays, the so called Hounsfield Units (HU). In the production of the treatment plan, one has to transform the map of HU into a map of RSP, since the patient is going to be treated with a beam of protons. However, there are unavoidable uncertainties associated with the derivation of the RSP map from the X-ray CT scan. Apart from the fact that the HU to RSP conversion depends on the chemical composition of the volume traversed by protons and not only on its HU value, it is not possible to ignore the ambiguity and limitations of the different HU to RSP conversion algorithms that are being used nowadays \cite{Wohlfahrt2020}. 

The aforementioned effects may lead to proton range uncertainties ranging from 2.1 to 2.2\%, depending on the organ: prostate, lung and head, when using dual-energy X-ray CT (DECT) \cite{Li2017}. These proton range uncertainties result in either a higher dose to healthy tissues or in a far too conservative treatment plan to avoid that. Reducing these uncertainties would allow a better planning that maximises the dose to the tumour, minimising at the same time the dose to the surrounding tissue. In order to reduce the uncertainty in proton range and take full advantage of the therapeutic potential of proton therapy, it is necessary to provide the treatment planning software with RSP maps obtained with proton beams rather than those derived after a conversion from the HU maps obtained with X rays. Proton computed tomography (proton CT) is the appropriate tool to produce such images since it makes use of proton beams provided by the same accelerator that is used later for the treatment, but this time at higher energy, so that the protons go through the patient and reach an appropriate proton scanner to form an image. See for instance the work of Takabe et al. \cite{Takabe2019}, for a descriptive introduction to proton CT. Some other recent studies with more advanced scanners are described in Dedes et al. \cite{Dedes2019} and Esposito et al. \cite{Esposito2018}. In the next section we will describe the basis of proton CT and our approach to a proton scanner for imaging.

Besides medical physics and imaging, basic nuclear-physics research in general involves the development of nuclear instrumentation, in the form of spectrometers and radiation detectors, to perform nuclear reactions and study the structure of matter. Within the detector R\&D process, the design and test of prototypes is  rather frequent. Sometimes, the prototypes are just small parts, the building blocks of the final device. In these cases, the optimised prototypes are often an integral part of the final product. However, some other times, the prototypes, although being very valuable radiation detectors themselves, cannot be used in the final device because they do not comply with the requirements or simply because they do not have the appropriate geometry: shape or size. 

In this work we will show how we have re-used one of these prototype detectors, that was developed as part of the R\&D of a larger device but cannot be used now as part of it. This, in combination with some other instrumentation used for nuclear reaction and structure experiments, has been converted into a proton scanner capable of performing proton radiography and tomography as will be described in the next sections.

\section{Materials and Methods}

\subsection{The proton CT scanner}
Any scanning technique based on the use of a non-invasive probe (e.g. penetrating radiation or ultrasounds), to obtain images of the inner body by sections, is known as tomography. These images by sections can be combined, using the appropriate reconstruction method, to form a 3-dimensional (3D) model of the object under study. In the case of transmission tomography, one generally starts by obtaining plane 2-dimensional (2D) projections, that, using the appropriate reconstruction algorithm, are turned into the final tomographic sections or 3D image. The most typical case of medical tomography technique by transmission is the X-ray CT, obtained from plane X-ray radiographs. The subject of this paper refers to tomography with proton beams, in other words, the reconstruction of images by sections using a proton beam as probe. Therefore, at the basis of proton CT stands the use of a proton accelerator that provides a proton beam with enough energy to go through the object of study. As an example, Johnson et al. \cite{Johnson2017} acquired proton images of head phantoms using 200-MeV proton beams. In clinical practice, the object of study is a part of a patient body. Since we are presenting here a pre-clinical instrument, from now on the object of study will be referred to as phantom. As for the case of X-ray CT, we will start by obtaining proton radiographs that will be useful by themselves as explained later.

\begin{figure}[h!]
\begin{center}
\includegraphics[width=8cm]{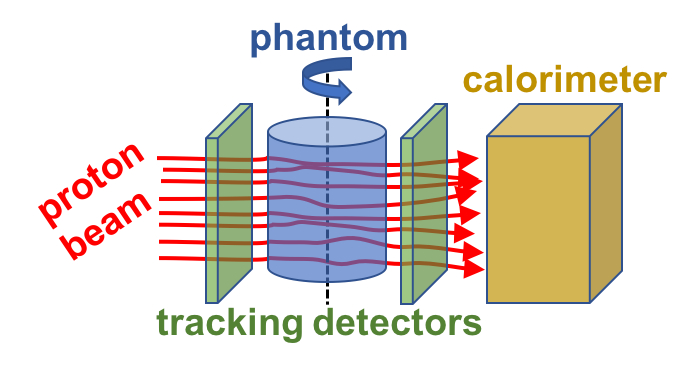}
\end{center}
\caption{Basic scheme of a proton-CT scanner. The proton beam (red) travels through the front tracking detector, the phantom, the rear tracking detector, and is fully stopped in the calorimeter (see text). \vspace*{5mm}}
\label{fig:basisCT}
\end{figure}

Since the main goal of the proton CT scan is to produce a map of RSP, we can either integrate the detected proton current, like in \cite{Rinaldi2013}, or detect the individual protons, as done in this work. We make use of tracking detectors to determine the trajectories of individual protons and a calorimeter, a detector that absorbs all the energy of the particles penetrating, to measure their residual energy. Fig.~\ref{fig:basisCT} shows a sketch of a simple proton-CT scanner. The tracking detectors, in green, are placed at the entrance and exit sides of the object of study, and are due to determine the entrance and exit point of each proton trajectory. In this work these trajectories are taken as straight lines as zero-order approximation, although it is known that there is always a certain deviation within the object mainly due to multiple Coulomb scattering. For this reason, the spatial resolution quoted in the following is just an upper limit. The use of more sophisticated reconstruction algorithms including curved trajectories like in \cite{Schulte2008} and refs. therein, would give a better spatial resolution, since our quoted value could be dominated by the reconstruction algorithm. Apart from the trajectory followed, to calculate the RSP of the different materials within the phantom, we need to know the energy deposited by the protons. Since we know the energy of the beam delivered by the accelerator, what we really want to measure is the residual energy of the protons after having passed through the object and tracking detectors. For that, we need to place a residual-energy detector, namely a calorimeter or a range telescope, right after the rear tracking detector.

For a concise description of the process to Fig.~\ref{fig:basisCT}. The proton beam reaches the setup, from the left side in the figure, and pass through the front position-sensitive detector, the phantom, and the rear position-sensitive detector. From the spatial coordinates of the two hits, the first one in the front detector and the second one in the rear detector, we can trace a straight line. This is the approximation to the trajectory taken in this work. After traversing the rear tracking detector, the protons leave their remaining kinetic energy in the bulk of the calorimeter, at the right-hand side in the figure. Combining the trajectories and residual energy measured for each proton, we can reconstruct tomographic images of the RSP in the bulk of the phantom following any of the methods detailed in \cite{Penfold2015}. \\

\begin{figure*}[h!]
\begin{center}
\includegraphics[width=0.85\textwidth]{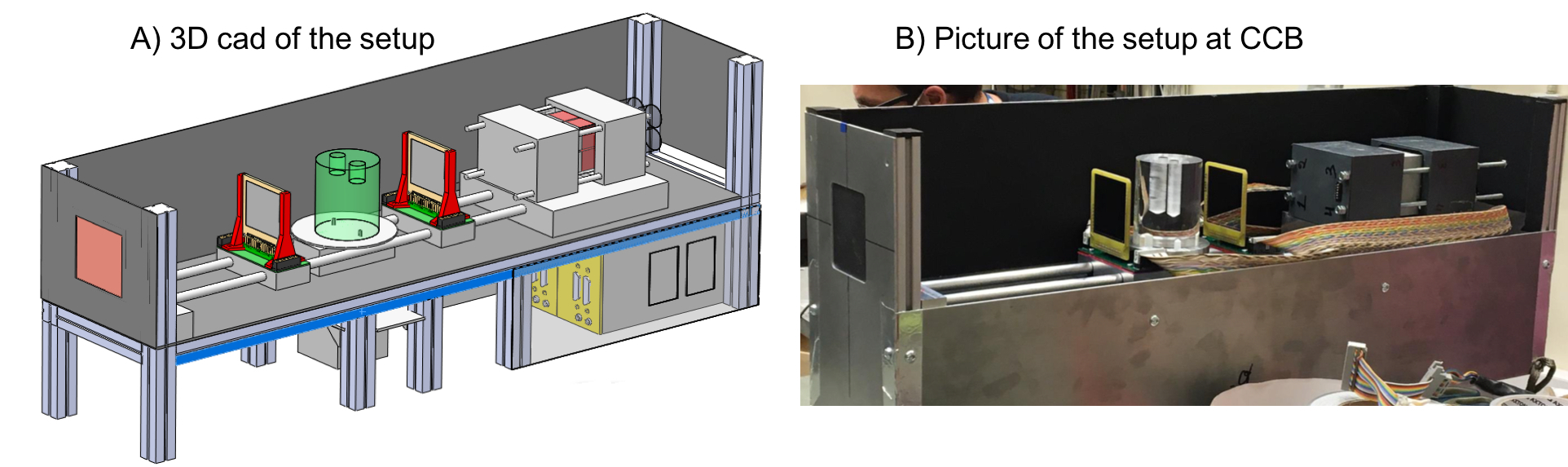}
\end{center}
\caption{The setup used is shown. On the left hand side, panel A shows the 3D-CAD drawing, while on the right hand side, panel B is a picture of the setup as used at CCB (Krakow).\vspace*{5mm}}
\label{fig:setup}
\end{figure*}

Fig. \ref{fig:setup} panel A shows a 3D-CAD design of our proton-CT scanner. In this sketch one can clearly see the front and rear tracking detectors held by their red supports, the green phantom cylinder between them, and the calorimeter at the right end of the setup. The full setup is enclosed in an opaque box to prevent the passage of light that would produce spurious signals in the tracking detectors. At the right-hand side, Fig. \ref{fig:setup} panel B shows a real picture of the actual setup. Details on the tracking and residual energy detectors are given in what follows.\\

{\it Double-Sided Silicon Detectors for proton tracking}

The tracking detector system is comprised of two Double-Sided Silicon Strip Detectors (DSSD), manufactured by Micron Semiconductor Ltd. The first DSSD detector is placed directly facing the proton beam, at the front side of the phantom, to determine the entrance point. The second one in placed at the rear position, to determine the exit point of the protons. Both DSSDs are 1-mm thick, and segmented into 16 vertical and 16 horizontal strips, giving a total of 256 pixels of 3x3 mm$^2$ per tracking detector. The two DSSD detectors where set 8 cm apart from each other, covering a field of view of 48$\times$48$\times$80 mm$^3$. A full description of these detectors and a very thorough characterisation of their response function to charged particles is given in \cite{Vinals2021}. During the measurements presented in this work, the signals from the DSSDs went through Mesytec preamplifiers and shapers before entering the CAEN V785 ADCs at the data acquisition system (ACQ).\\

{\it CEPA4: The Residual-Energy Detector}

\begin{figure}[h!]
\begin{center}
\includegraphics[width=0.5\textwidth]{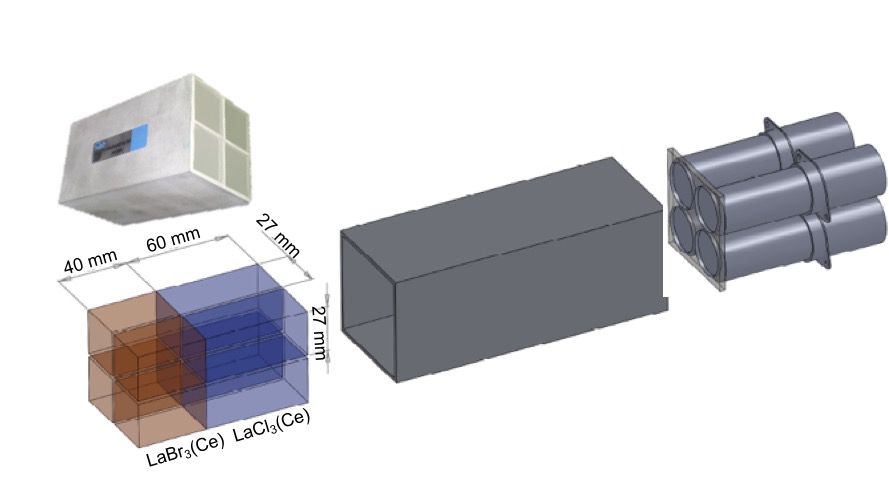}
\end{center}
\caption{CEPA4 detector assembly. At the left-hand side, the 4 double crystals of 4 cm of LaBr$_3$(Ce) and 6 cm of LaCl$_3$(Ce) are seen in schematic mode and in a picture of the real assembly. The middle part represents the 1 mm Aluminum canning and, at the right-hand side, the 4 photomultiplier tubes that collect the scintillator light.\vspace*{5mm}}
\label{fig:CEPA4}
\end{figure}

The calorimeter, or residual-energy detector, used in our scanner is an array of four scintillation units, each of them comprised of two scintillator crystals in phoswich configuration: 4 cm of LaBr$_3$(Ce) and 6 cm of LaCl$_3$(Ce) with a common photomultiplier tube (see Fig. \ref{fig:CEPA4}). The crystals are individually wrapped in reflecting material and closed packed in a 0.5 mm Aluminum can. The full detector array, called CEPA4, is a prototype detector for the endcap of CALIFA, the electromagnetic calorimeter of R$^3$B at FAIR. A full description of the CEPA4 and its response to high-energy proton beams can be found in \cite{Nacher2015}. In all the measurements described in this work, the signals from the photomultipliers were directly acquired by a Mesytec MDPP-16-QDC high-resolution time and charge integrating digitizer at the DAQ. The main advantage of using CEPA4 as residual-energy detector lies in its energy resolution, that translates in a better contrast in the final RSP image. For protons of 80-130 MeV, which are the relevant energies for our study, the protons are stopped in the first crystal, namely the LaBr$_3$(Ce) part of the phoswich. In this crystal the energy resolution ranges within 2 to 3$\%$.

\subsection{In-beam experiments}
\label{sec:exp}

Apart from the setup and fine-tuning of the system at the laboratory of Instituto de Estructura de la Materia (IEM-CSIC, Madrid), we have carried out two experiments with proton beams: one at Centro de Microan\'alisis de Materiales (CMAM, Madrid), and the other at the Centrum Cyklotronowe Bronowice (CCB, Krakow). The former, at CMAM, was a proof-of-concept experiment with low-energy proton beams to test the DSSD as tracking detectors, details on the results can be found in \cite{Briz2021}. The latter, at CCB, was the first test of the full setup with high-energy protons and is detailed in what follows.

For a realistic test of our proton-CT scanner we used the high-energy proton beams provided by the IBA PROTEUS C235 proton cyclotron at the CCB in Krakow. The latter is part of the Henryk Niewodnicza\'nski  Institute of Nuclear Physics Polish Academy of Sciences in Krakow (IFJ PAN) and its main focus is the application of cyclotrons in scientific research and tumor radiotherapy. For our measurements we were provided with mono-energetic proton beams at energies 100 and 110 MeV, with an energy spread of 1.5\% (FWHM).

\begin{figure*}[h!]
\begin{center}
\includegraphics[width=0.9\textwidth]{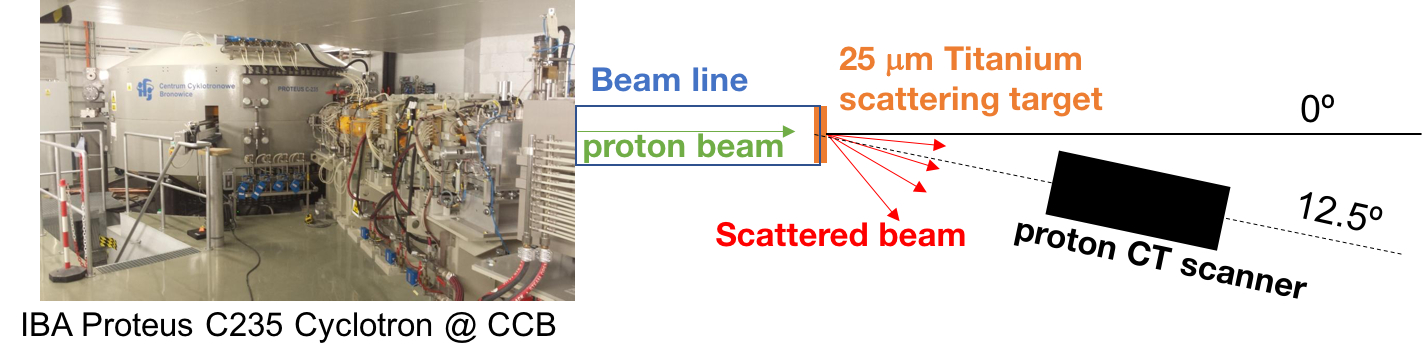}
\end{center}
\caption{Experimental setup used in the CCB experiment in Krakow (Poland). Left panel: a picture of the Proteus C235 cyclotron and the beam line to the experimental hall. Right panel: sketch of the setup at the experimental hall, where the proton beams of 100-110 MeV were scattered in a 25-$\mu$m-thick Titanium target. Our proton CT scanner, placed at 12.5$^{\circ}$ with respect to the incident beam direction, was receiving the scattered proton beam.\vspace*{5mm}}
\label{fig:expCCB}
\end{figure*}

The accelerator provided a high-current pencil beam ($\approx$ 1 nA, $\approx$ 10 mm diameter), however, for our purposes, we needed a low-current fan beam covering the full field of view of our scanner. Thus, we measured using the protons scattered on a 25-$\mu$m thick (11.25 mg/cm$^2$) Titanium foil. The measurement was performed in air and with the proton-CT scanner at an angle of 12.5 degrees with respect to the beam direction, the alignment of the system was done using a laser system. In these conditions, our acquisition rate was 700 Hz of triple coincidences (with a total rate of 10 kHz in OR trigger condition). We kept such a low counting rate since the DAQ was in use for the first time and we wanted to avoid pile-up and dead-time effects as much as possible. However, our current tests at the lab indicate that we can run experiments with triple-coincidence conditions withstanding counting rates up to 45 kHz with no significant pileup and less than 10\% of dead-time. The energy loss due to the scattering angle, and the losses in the Ti foil have been calculated using the \emph{\textsc{GEANT4}} Monte Carlo code \cite{Allison2016}, the losses in the DSSD tracking detectors have been directly acquired by the detectors themselves since they also perform well as spectrometers. For a uniform phantom (PMMA cylinder of 6-cm diameter) and 100-MeV protons, the measured energy loss in the front and rear tracking detectors (1-mm thick) were 1.4 and 2.8 MeV, respectively. We used proton beams of 95, 100 and 120 MeV to calibrate the tracking and residual energy detectors, but the final measurements for radiography and tomography where carried out at 100 and 110 MeV respectively. For more details on the calibration procedure the reader is referred to Ref. \cite{Briz2022}. A detailed Monte Carlo simulation of the experiment was performed using \emph{\textsc{GEANT4}} \cite{Allison2016} to calculate the values of energy deposited in the different volumes for the three proton beam energies, which allowed for an accurate calibration in the energy range of interest. A picture of the cyclotron providing the proton beam at CCB and a schematic view of the setup are shown in Fig. \ref{fig:expCCB}. 
\\

For our measurements at CCB we used some custom-made phantoms specially designed to test the performance of the system in terms of spatial resolution and RSP estimation. For the case of radiography phantoms, we enclosed 10-mm-thick Aluminum inserts into a 50-mm-thick PMMA matrix. We tested two different patterns: a cross and a point/line regular spatial pattern. A picture of these two Aluminum inserts included in the two phantoms can be seen in Fig. \ref{fig:radiographs}, left column. The quality of the images was studied via a Modulation Transfer Function (MTF) analysis using the profiles obtained with the regular spatial pattern (holes in C and D panels in Fig. \ref{fig:radiographs}). The MTF is a measure of the capability of our device to transfer contrast at a particular resolution from the object to the image. In other words, the MTF is a way to incorporate resolution and contrast into a single specification. In this study, the MTF is calculated as the contrast (in percentage of grey level) in the image between one hole and the Aluminum spacing, and it is represented, as a function of the number of line pairs (hole-spacing pairs in our case) per mm as will be shown in the next section.\\

For the case of tomography, we designed two different phantoms based on PMMA cylinders. The first one was a Derenzo-like pattern with holes of 7, 5 and 3 mm diameter and with separations of the same length. In order to take several projections at different angular positions, the phantom was placed on a rotatory platform connected to a step motor. The measurements were carried out at a proton energy of 110 MeV. To test the RSP mapping capabilities of our setup we designed a second phantom: a PMMA cylinder of 60-mm diameter with two inserts of 9-mm diameter each that can be filled with different liquids, gels or powders. Since the commercial PMMA materials may have small differences in their density, we have determined the density by precisely measuring the dimensions of the phantom ($\pm$ 0.01 mm) and its weight ($\pm$ 0.02 g). Pictures of the cylindrical phantoms used in the tomography tests are shown in the figures presented in the next section.

\section{Results}

\subsection{Proton Radiography}

As we explained before, in the measuring process for proton CT we will obtain proton radiographs, 2D images that are useful by themselves. While the slices from the tomographic reconstruction hold a direct measurement of the RSP, the plane radiographs hold information of the line integrals of the RSP. This line integrals of the RSP are referred to as Water Equivalent Path Length (WEPL), and when they are averaged within a certain spatial bin, they turn into the so-called water-equivalent thickness (WET). Proton radiographs, when the spatial resolution allows for it, can be used for patient alignment/positioning. Furthermore, a comparison between a real proton radiograph and a virtual proton radiograph reconstructed from the X-ray CT used for the treatment plan, can be a very powerful tool to detect possible proton range errors due to the conversion of HU to RSP before the treatment.

\begin{figure}[h!]
\begin{center}
\includegraphics[width=8cm]{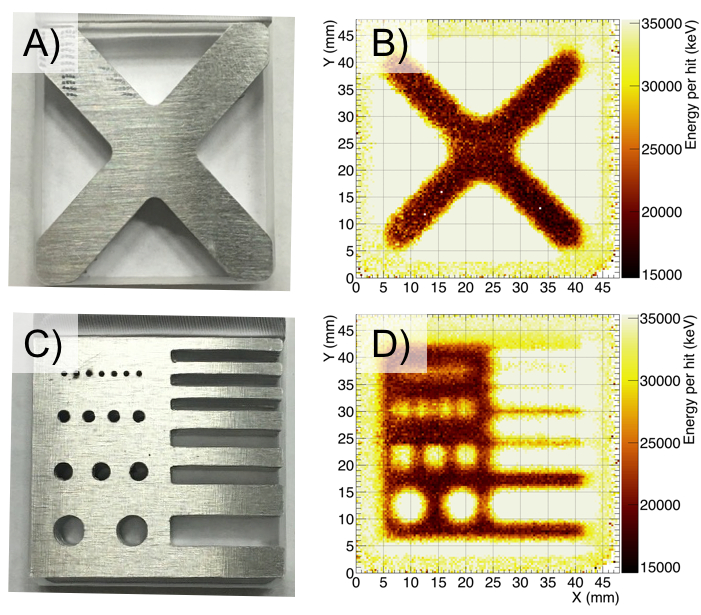}
\end{center}
\caption{Radiographs obtained using a 100-MeV proton beam at CCB facility in Krakow (Poland) with our proton-CT scanner. The left panels A and C show pictures of the phantoms used for the radiographic study. The obtained transmission images are displayed in panels B and D. Panel D shows a regular spatial pattern that was used to perform the MTF analysis mentioned in the text and described in detail in \cite{Briz2022}. The dots at the 3rd raw from the bottom have 2 mm diameter and are 2 mm apart from each other and they are well resolved in the proton image.\vspace*{5mm}}
\label{fig:radiographs}
\end{figure}

The phantoms used to test the performance of the scanner in radiography mode have been described in the previous section, and the results of the proton radiographs obtained with our scanner with proton beams of 100 MeV can be seen in Fig. \ref{fig:radiographs}, right column. The radiographs reconstructed in the figure were obtained at the central plane of the phantom. The X,Y coordinates were determined by simply averaging the X and Y coordinates at both detector planes, always assuming straight proton trajectories. The colour scale represents the average energy deposited per detected proton. It is important to recall here that the energy deposited in the phantom is not proportional to the RSP but to its line integral along the proton trajectory and averaged within a spatial bin, namely the WET. For a rough estimation of the spatial resolution one can look carefully at the bottom half of Fig. \ref{fig:radiographs}. In the picture of the phantom, displayed in C, at the left hand side, the holes at the third raw starting from the bottom have 2 mm of diameter and are 2 mm apart of each other. In the radiograph, at the right hand side, one can clearly see that these holes are well resolved in the image, however, the 1-mm holes at the upper row are not. This allows us to conclude that the spatial resolution is better than 2 mm.

\begin{figure}[h!]
\begin{center}
\includegraphics[width=8cm]{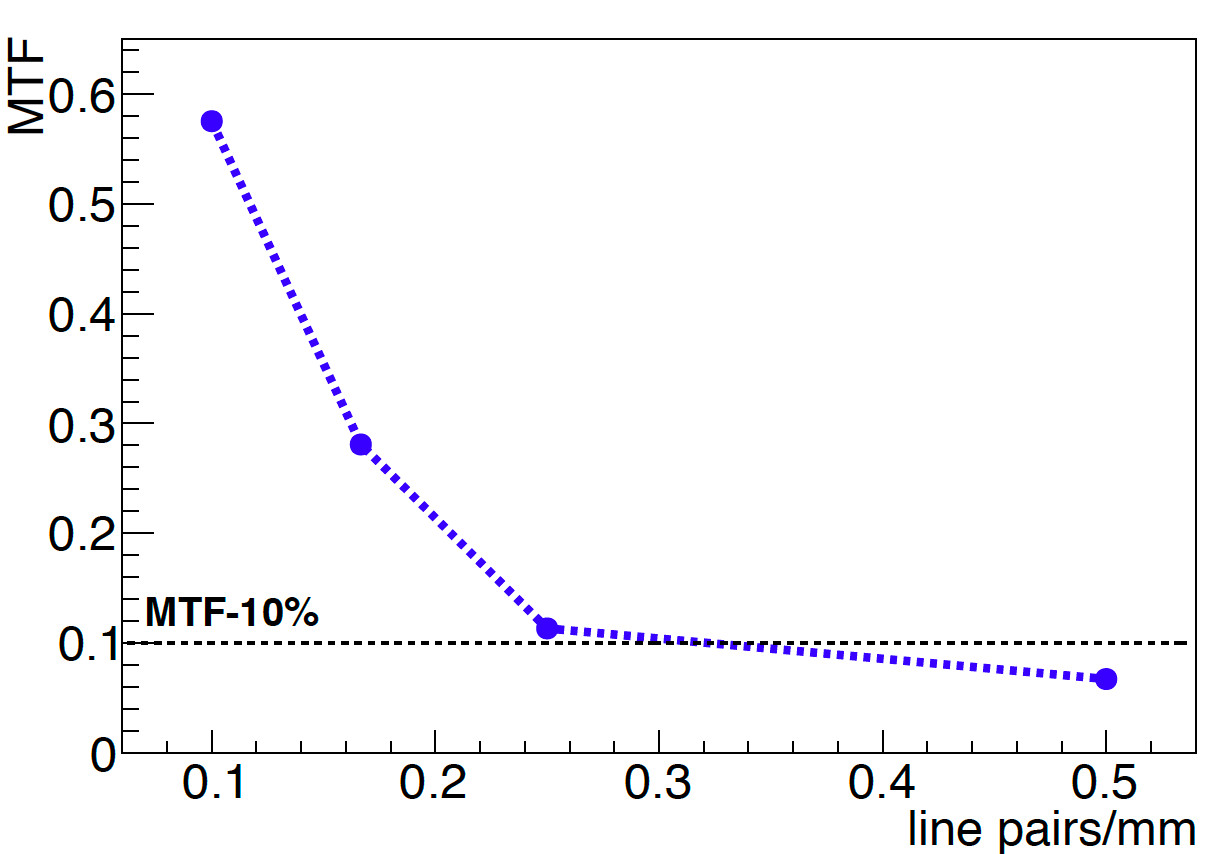}
\end{center}
\caption{Modulation Transfer Function (MTF) study of the line profiles corresponding to the region with holes in Fig. \ref{fig:radiographs}. The reference value of MTF=10$\%$ is indicated with a horizontal dashed line.\vspace*{5mm}}
\label{fig:MTF}
\end{figure}

A detailed description of the radiography measurements and results has already been published in \cite{Briz2022}. The quality of the images was studied via the MTF analysis explained before. Fig. \ref{fig:MTF} represents the MTF as a function of the number of line pairs (hole-spacing pairs in our case) per mm. Looking at the resolved lines in Fig. \ref{fig:radiographs}, we conclude that the spatial resolution of the device is better than 2 mm. The MTF-10\% = 0.3 line-pairs/mm, deduced from Fig. \ref{fig:MTF}, corroborates this 2-mm upper limit for the spatial resolution and is comparable to those of other existing devices (e.g. MTF-10\% = 0.6 line-pairs/mm in \cite{Sarosiek2021}). For more details the reader is referred to \cite{Briz2022}.
\\

\subsection{Proton Tomography}
For the purpose of this work, with emphasis in the validation of the device as proton-CT scanner, we will only present images obtained with a simple filtered back-projection (FBP) algorithm, using a ramp filter, that assumes straight paths for the protons inside the phantom. With this approach, we have performed two different measurements, one to work on the different tomographic acquisition parameters and estimate the spatial resolution of the system, and another one to check its capability to resolve the proton RSP values of different materials. For these measurements we designed two different phantoms based on PMMA cylinders and described in the previous section.

\begin{figure*}[h!]
\begin{center}
\includegraphics[width=0.8\textwidth]{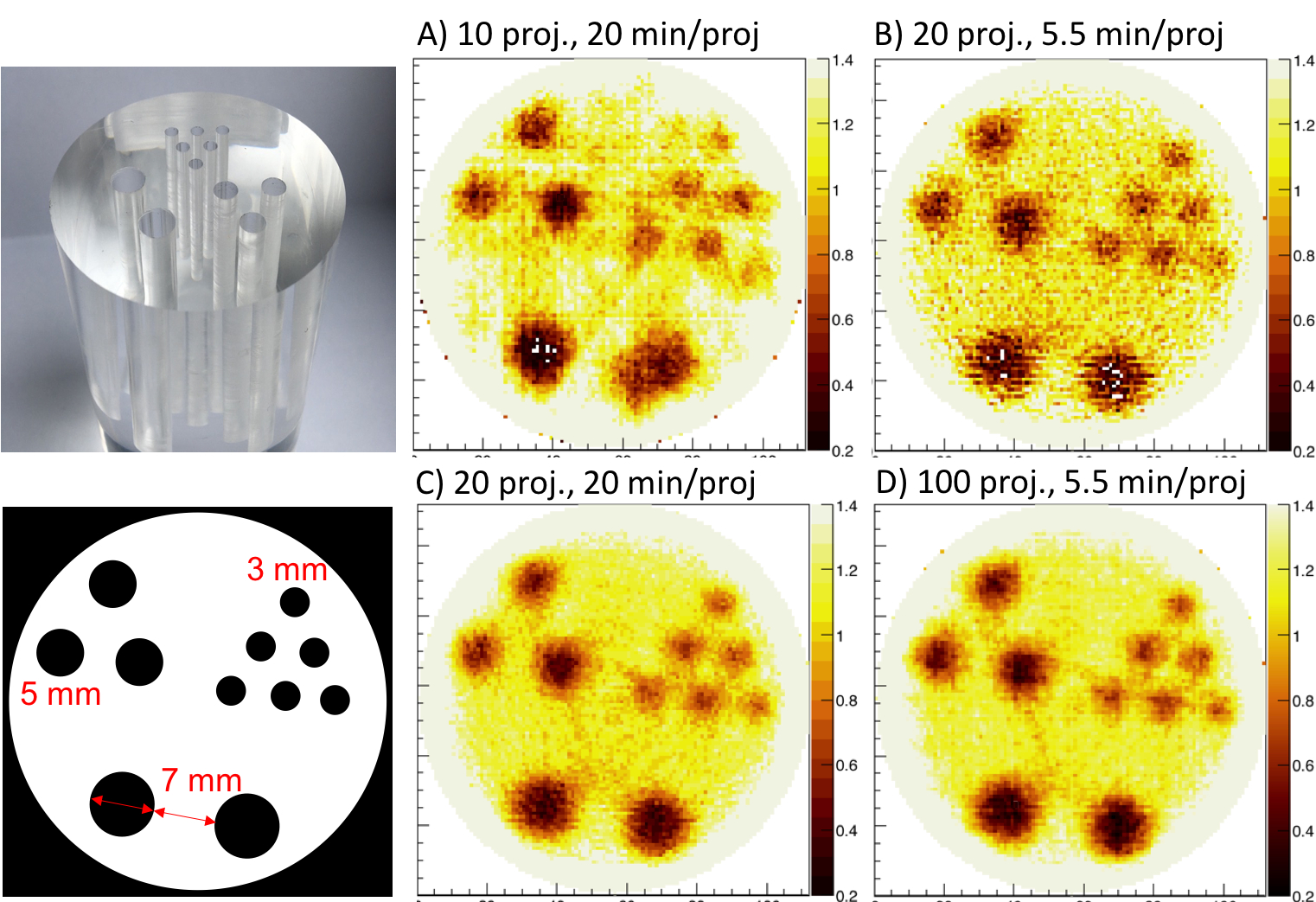}
\end{center}
\caption{Derenzo-like phantom and pattern with holes and separations of 3, 5 and 7 mm and the corresponding 4 images for different scans performed. The pixel size is 0.4 $\times$ 0.4 mm$^2$ in the 4 images. A) The scan was done with 10 projections (20 min each) in steps of 18$^{\circ}$, B) 20 projections (5.5 min each) in steps of 9$^{\circ}$, C) 20 projections (20 min each) in steps of 9$^{\circ}$ D) 100 projections (5.5 min each) in steps of 1.8$^{\circ}$.\vspace*{5mm}}
\label{fig:ct1}
\end{figure*}

The Derenzo-like phantom is shown on leftmost panel of Fig. \ref{fig:ct1}. The A, B, C and D panels of the same figure show the filtered back-projected images, using pixels of 0.4 $\times$ 0.4 mm$^2$ in the image plane, under different measurement conditions: A) 10 projections of 20 minutes each, in steps of 18$^{\circ}$; B) 20 projections of 5.5 minutes each, in steps of 9$^{\circ}$; C) 20 projections of 20 minutes each, in steps of 9$^{\circ}$; D) 100 projections of 5.5 minutes each, in steps of 1.8$^{\circ}$. The total number of projections of each measurement shown in the figure always cover half a turn, i.e., 180$^{\circ}$. During these measurements the proton current was stable at around 1 nA and, at this intensity, we counted $\approx$700 triple coincidences per second (front DSSD and rear DSSD and Calorimeter). In these conditions, the projections of 5.5 minutes recorded $\approx$2.3$\times 10^5$ events, whereas the projections of 20 minutes recorded $\approx$8.4$\times 10^5$. The difference in statistics per projection, as well as the different number of projections affect considerably the image quality. Looking at the four images of Fig. \ref{fig:ct1} we can clearly appreciate, firstly, that the image with the lowest number of projections, panel A), does not reproduce fairly the pattern, since one of the cylinders of 7 mm has not the shape of a cylinder and two of the cylinders of 3 mm are blurred and practically absent.  Secondly, panel B) shows the image with low statistics per projection (5.5 min) but 20 projections in total, and it already reproduces fairly well the pattern, since all cylinders are seen with the right shape and position. Going from A) to B) shows that the effect of lowering the statistics per projection is well compensated by taking a higher number of projections. The third panel C) keeps the same number of projections than B) but increasing the statistics per projection and the improvement is obvious. Finally, we took a longer measurement of 100 projections of lower statistics that is shown in D). In this case the result is more uniform, but we do not see a better resolution than in the previous image, indicating that, with a proper measurement of a uniform cylinder for normalization, 20 projections covering 180$^{\circ}$ is an acceptable sampling for our purposes. The main limitation in statistics/time in this study was due to the high dead-time of the DAQ. However, recently we have carried out a new series of measurements with the same system but an improved electronic setup and digitization configuration, being able to take similar images with less than 10$\%$ of dead-time at counting rates of 45 kHz of triple coincidences. This means a factor 64 improvement with respect to the conditions we had at CCB when we performed the measurements shown here. In the current conditions our device would take an image at the contrast and quality level achieved in Fig. \ref{fig:ct1} C) in 6.3 min and D) in 8.6 min, what compares better to other similar devices in the field (see Table 1 of Ref. \cite{Johnson2018} for a complete list).\\

\begin{figure*}[h!]
\begin{center}
\includegraphics[width=0.8\textwidth]{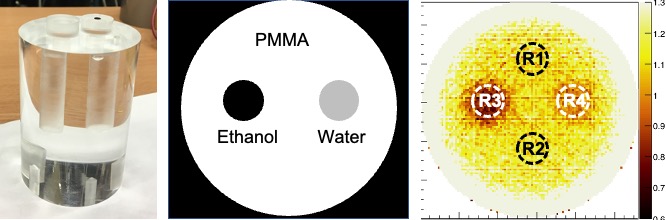}
\end{center}
\caption{Phantom of PMMA with two inserts of 9-mm diameter filled with with ethanol (purity$>$99.9 \% \cite{Eth22}) and distilled water, used to evaluate the capabilities of our device for RSP determination. The resulting image is shown at the right panel, where the four different regions of interest considered are indicated with black/white circles (6-mm diameter) and letters. Those regions were used to determine the values of RSP for ethanol and PMMA indicated in Table 1, obtained after normalizing the image with respect to the water region R4.\vspace*{5mm}}
\label{fig:ct2}
\end{figure*}

Beyond the capabilities to produce tomographic images with resolving power better than 3 mm, our main goal in this work is to produce reliable RSP maps. In this context, the energy resolution of the residual-energy detector is crucial, since the energy deposited by the protons in the traversed volumes depends completely on the RSP of the material. This is why our proton-CT scanner, even being made of detectors that were originally designed for other use, is very promising in terms of RSP mapping, since the residual-energy detector is made of high-resolution scintillators. To test the RSP mapping capabilities of our setup we designed a special phantom, a PMMA cylinder of 60-mm diameter with two inserts of 9-mm diameter each that can be filled with different liquids, gels or powders. A picture of such phantom is shown at the leftmost panel of Fig. \ref{fig:ct2}.
\\

We performed proton scans at 110 MeV of the phantom with the inserts filled with ethanol (purity$>$99.9 \% \cite{Eth22}) and distilled water. We took 10 projections of 20 minutes each, in steps of 18$^{\circ}$, covering 180$^{\circ}$ in total. As with the previous scans of the Derenzo-like phantom, we have used a simple filtered back-projection with the ramp filter to reconstruct the images. \\

The rightmost panel of Fig. \ref{fig:ct2} shows the four regions of interest (ROIs) that have been defined to study each material present in our phantom, one ROI of water, one of ethanol and two ROIs covering the PMMA matrix. The reconstructed image was consequently normalised to water in order to estimate the RSP of PMMA and ethanol. The resulting values are shown in Table 1, where they are compared with the experimental values reported in Ref. \cite{Abbema2018}, that were measured using proton beams of 149 MeV. The two first columns of the table show the densities of the ethanol and PMMA from Ref. \cite{Abbema2018} and the ones from this work. The similarity in material densities validates the comparison. The third and fourth columns show the values and uncertainties of the RSP of PMMA and ethanol that have been obtained, after the normalisation of the image with respect to the region of water (R4 in Fig. \ref{fig:ct2}), as the mean RSP value and the standard error of the mean, obtained inside the respective ROIs indicated in Fig. \ref{fig:ct2} as R1 and R2 for PMMA and R3 for ethanol. These ROIs have been taken as circumferences of diameter 6 mm (to be compared to 9-mm diameter for the actual inserts) for the calculation of the RSP. The latter has been calculated as the mean value. The last column of Table 1 shows the relative difference between the present RSP values and those taken from \cite{Abbema2018} as reference values, being in both cases of the order of 1$\%$. It can be noticed that the deviations of densities and RSPs follow the same tendency.
\\
\begin{table*}[ht]
\caption{Comparison of densities and RSP values obtained in this work, using 110-MeV protons, with the values obtained in Ref. \cite{Abbema2018}, using 149-MeV protons. The RSP values from this work were obtained as the average of the RSP values of the pixels included in the ROIs indicated in Fig. \ref{fig:ct2} after normalizing the full image with the region of water. The uncertainties in brackets have been calculated as standard error of the mean value within the ROI.}
\begin{tabular}{c c c c c c}
\\
\hline 
\rule{0pt}{2ex}   \bf{Region} &\bf{Density (Ref. \cite{Abbema2018})}& \bf{Density (this work)} & \bf{RSP (Ref. \cite{Abbema2018})} & \bf{RSP (this work)} & \bf{Relative difference}\\  [1ex]
\hline\hline
\rule{0pt}{3ex}   {\bf Water}   &N/A &N/A &  1.000(4)       & 1.000(5)   & N/A \\  [1ex]
\hline 
\rule{0pt}{3ex}   {\bf Ethanol} &0.788(1) g/cm$^3$& 0.790(1) g/cm$^3$ &0.822(3)      & 0.829(7)  & +0.85 $\%$\\  [1ex]
\hline 
\rule{0pt}{3ex}   {\bf PMMA}    &1.183(4) g/cm$^3$& 1.179(14) g/cm$^3$ &1.168(5)     & 1.158(5) &  -0.86$\%$ \\  [1ex]
\hline
\vspace*{5mm}
\end{tabular}
\end{table*}

Our results and those of Ref. \cite{Abbema2018} are in agreement within the uncertainties. The resulting proton RSP  map from our test beam is satisfactory for a first experiment. However, the relative differences are not negligible, definitely a bit worse than those reported e.g. in the recent work of Dedes et al. \cite{Dedes2019}, and the relative error on our values are 8$\%$ for Ethanol and 4$\%$ for PMMA, far worse than those of \cite{Dedes2019}. We remind here the reader that this was the first test of this proton-CT scanner that is made of detectors originally designed for different applications. This RSP map was obtained only with 10 projections and without any uniformity correction. Taking the optimal 20 projections and correcting the data with a dedicated measurement of a uniform PMMA phantom will decrease the uncertainties and improve both the spatial and RSP resolution and accuracy.\\

Data sets generated during the current study are available from the corresponding author on reasonable request.

\section{Discussion}

The results shown in the previous section validate our setup as a prototype of proton-CT scanner. The spatial resolution of our setup has room for improvement, for instance there are DSSDs in the market with much higher granularity. We have experimentally determined a spatial resolution between 1 and 2 mm  in radiography, but this is highly dependent and limited by the pixel size of our DSSD detectors (3 mm x 3 mm). However, for a proof of concept we have demonstrated that our scanner can resolve 2 mm holes in radiography and 3 mm in tomography images. Despite using a simple approximation to the reconstruction problem including only straight-line paths for the protons and an analytic FBP reconstruction algorithm, we reached to solve structures as small as 3 mm in tomography. We have to take this number as an upper limit for two reasons: firstly, the image reconstruction would be better if we could include curved paths for the protons and, secondly, since we did not expect to resolve 3 mm we did not build a Derenzo-like cylinder with smaller holes to really find the limit. The excellent energy resolution of the scintillator crystal (LaBr$_3$) used for the residual-energy detector, allows for a fairly good resolving power in RSP in the tomographic images. The RSP of ethanol and PMMA materials have been reproduced accurately. Additional tests will be performed to deeper evaluate the potential of our first prototype in tomography and to reduce the relative uncertainties in our reconstructed RSP values.

The main concern during the first set of measurements presented here was the dead-time of the acquisition system that only allowed for measurements at low counting rates (700 Hz of triple coincidences), meaning very long scanning times. This would be a showstopper for the future of our device as proton-CT scanner, however, recently we have optimised our electronics and DAQ and carried out a new set of measurements with different phantoms. With some improvements at the digitisation level, we have been able to take data with less than 10$\%$ of dead-time at counting rates of 45 kHz of triple coincidences, closer to other pre-clinical scanners. This translates into a much faster system capable to take the images presented here in few minutes rather than hours, and compares well to other similar devices in the field.

\begin{acknowledgements} 
This work has been mainly supported by the PRONTO-CM and ASAP-CM projects ( grants B2017/BMD-3888 and P2022/BMD-7434 fromComunidad de Madrid, Spain) that has sponsored J.A. Briz and A.N. Nerio. The experiments have been carried out with the support of the European Union Horizon 2020 research and innovation programme under grant agreement no. 654002 (ENSAR2) and grant agreement No [730983] (INSPIRE). This publication is also part of the R\&D grants PID2022-140162NB-I00, PID2022-138297NB-C21 and PDC2022-133382-I00, funded by the Spanish Ministry of Science (MCIN/AEI/10.13039/501100011033) and grant CIPROM/2021/064 from Generalitat Valenciana. The authors want to express their gratefulness to the CCB crew for their unconditional help during the data taking.
\end{acknowledgements}

\bibliographystyle{spphys}       
\bibliography{pCTbibliography}   

\begin{thebibliography}{10}
\providecommand{\url}[1]{{#1}}
\providecommand{\urlprefix}{URL }
\expandafter\ifx\csname urlstyle\endcsname\relax
  \providecommand{\doi}[1]{DOI \discretionary{}{}{}#1}\else
  \providecommand{\doi}{DOI \discretionary{}{}{}\begingroup
  \urlstyle{rm}\Url}\fi

\bibitem{Wohlfahrt2020}
P.~Wohlfahrt, C.~Richter, Br. J. Radiol. \textbf{93}, 20190590 (2020).
\newblock \doi{10.1259/bjr.20190590}

\bibitem{Li2017}
B.~Li, H.C. Lee, X.~Duan, et~al., Phys. Med. Biol. \textbf{62}, 7056 (2017).
\newblock \doi{10.1088/1361-6560/aa7dc9}

\bibitem{Takabe2019}
M.~Takabe, T.~Masuda, M.~Arimoto, et~al., Nucl. Inst. and Methods A
  \textbf{924}, 332 (2019).
\newblock \doi{10.1016/j.nima.2018.05.034}

\bibitem{Dedes2019}
G.~Dedes, J.~Dickmann, K.~Niepel, et~al., Phys. Med. Biol. \textbf{64}, 165002
  (2019).
\newblock \doi{10.1088/1361-6560/ab2b72}

\bibitem{Esposito2018}
M.~Esposito, C.~Waltham, et~al., Physica Medica \textbf{55,}, 149 (2018).
\newblock \doi{10.1016/j.ejmp.2018.10.020}

\bibitem{Johnson2017}
R.P. Johnson, V.~Bashkirov, G.~Coutrakon, et~al., Physics Procedia \textbf{90},
  209 (2017).
\newblock \doi{10.1016/j.phpro.2017.09.060}

\bibitem{Rinaldi2013}
I.~Rinaldi, S.~Brons, J.~Gordon, et~al., Phys. Med. Biol. \textbf{58}, 413
  (2013).
\newblock \doi{10.1088/0031-9155/58/3/413}

\bibitem{Schulte2008}
R.~Schulte, S.~Penfold, J.T. Tafas, et~al., Med. Phys. \textbf{35}, 4849
  (2008).
\newblock \doi{10.1118/1.2986139}

\bibitem{Penfold2015}
S.~Penfold, Y.~Censor, Sensing and Imaging \textbf{16}(19) (2015).
\newblock \doi{10.1007/s11220-015-0122-3}

\bibitem{Vinals2021}
S.~Vi{\~n}als, E.~N\'acher, O.~Tengblad, et~al., Eur. Phys. J. A
  \textbf{57}(2), 49 (2021).
\newblock \doi{10.1140/epja/s10050-021-00371-5}

\bibitem{Nacher2015}
E.~N\'acher, M.~M{\aa}rtensson, O.~Tengblad, et~al., Nucl. Instr. and Methods A
  \textbf{769}, 105 (2015).
\newblock \doi{10.1016/j.nima.2014.09.067}

\bibitem{Briz2021}
J.A. Briz, I.~Posadillo, V.G. T\'avora, et~al., EPJ Web of Conferences
  \textbf{253}, 09008 (2021).
\newblock \doi{10.1051/epjconf/202125309008}

\bibitem{Allison2016}
J.~Allison, K.~Amako, J.~Apostolakis, et~al., Nucl. Instr. and Methods A
  \textbf{835}, 186 (2016).
\newblock \doi{10.1016/j.nima.2016.06.125}

\bibitem{Briz2022}
J.A. Briz, A.N. Nerio, C.~Ballesteros, et~al., IEEE Transactions on Nuclear
  Science \textbf{69}, 696 (2022).
\newblock \doi{10.1109/TNS.2022.3142618}

\bibitem{Sarosiek2021}
C.~Sarosiek, E.A. DeJongh, G.~Coutrakon, et~al., Med. Phys. \textbf{48}, 2271
  (2021).
\newblock \doi{10.1002/mp.14801}

\bibitem{Johnson2018}
R.P. Johnson, Rep. Prog. Phys. \textbf{81}, 016701 (2018).
\newblock \doi{10.1088/1361-6633/aa8b1d}

\bibitem{Eth22}
J.t.baker absolute ethanol.
\newblock \urlprefix\url{https://es.vwr.com/store/product/11833037/null}

\bibitem{Abbema2018}
J.K. van Abbema, M.J. van Goethem, J.~Mulder, et~al., Nucl. Instr. and Methods
  B \textbf{436}, 99 (2018).
\newblock \doi{10.1016/j.nimb.2018.09.015}

\end{thebibliography}



\end{document}